\documentclass[conference]{IEEEtran}
\def\BibTeX{{\rm B\kern-.05em{\sc i\kern-.025em b}\kern-.08em
    T\kern-.1667em\lower.7ex\hbox{E}\kern-.125emX}}

\usepackage{url} 
\usepackage{booktabs}
\usepackage{colortbl}
\usepackage{xcolor}
\usepackage{multirow}
\usepackage{makecell}
\usepackage{amsmath}
\usepackage{amssymb}
\usepackage{graphicx}
\usepackage{textcomp}
\usepackage{stmaryrd}
\usepackage{algorithm}
\usepackage{algpseudocode}
\usepackage{mathtools}
\usepackage{parcolumns}
\usepackage{listings}
\usepackage{parcolumns}
\usepackage{xspace}
\usepackage{placeins}
\usepackage[T1]{fontenc}
\usepackage{tcolorbox}
\tcbuselibrary{skins, raster}

\def\smtencode#1{\llbracket #1 \rrbracket}
\def\mcount#1{|#1|}

\def\A{\mathbb{A}}
\def\R{\mathbb{R}}
\def\P{\mathbb{P}}

\def\Pallow{\mathbb{P}_{\mathit{Allow}}}
\def\Pdeny{\mathbb{P}_{\mathit{Deny}}}
\def\Pone{\mathbb{P}_1}
\def\Ptwo{\mathbb{P}_2}
\def\smtP{\smtencode{\P}}

\def\smtPone{\smtencode{\Pone}}
\def\smtPtwo{\smtencode{\Ptwo}}

\def\AllowP{\textsc{Allow}(\mathbb{P})}

\def\actionnamespace#1{\textsc{#1}\xspace}
\def\ec2{\actionnamespace{ec2}}

\def\s3{\actionnamespace{s3}}
\def\quacky{\actionnamespace{quacky}}
\def\PolicySummarizer{\actionnamespace{PolicySummarizer}}
\def\tool{\PolicySummarizer}

\newcommand{\rev}[1]{#1}

\colorlet{punct}{red!60!black}
\definecolor{background}{HTML}{EEEEEE}
\definecolor{delim}{RGB}{20,105,176}
\colorlet{numb}{magenta!60!black}
\usepackage{booktabs}
\usepackage{multirow}
\usepackage{makecell}
\lstdefinelanguage{json}{
    basicstyle=\scriptsize\ttfamily,
    showlines=true,
    frame=tlrb,
    backgroundcolor=\color{background},
    literate=
     *{0}{{{\color{numb}0}}}{1}
      {1}{{{\color{numb}1}}}{1}
      {2}{{{\color{numb}2}}}{1}
      {3}{{{\color{numb}3}}}{1}
      {4}{{{\color{numb}4}}}{1}
      {5}{{{\color{numb}5}}}{1}
      {6}{{{\color{numb}6}}}{1}
      {7}{{{\color{numb}7}}}{1}
      {8}{{{\color{numb}8}}}{1}
      {9}{{{\color{numb}9}}}{1}
      {:}{{{\color{punct}{:}}}}{1}
      {,}{{{\color{punct}{,}}}}{1}
      {\{}{{{\color{delim}{\{}}}}{1}
      {\}}{{{\color{delim}{\}}}}}{1}
      {[}{{{\color{delim}{[}}}}{1}
      {]}{{{\color{delim}{]}}}}{1},
}

\begin{document}
\title{Neurosymbolic Characterization for Reliable Access Control Policy Analysis}

\author{\IEEEauthorblockN{Adarsh Vatsa, Bethel Hall, and William Eiers}
\IEEEauthorblockA{Stevens Institute of Technology}}

\maketitle

\begingroup
\renewcommand\thefootnote{}
\footnotetext{Accepted to the IEEE International Symposium on Software Reliability Engineering (ISSRE 2026). \copyright~2026 IEEE. Personal use of this material is permitted. Permission from IEEE must be obtained for all other uses, in any current or future media, including reprinting/republishing this material for advertising or promotional purposes, creating new collective works, for resale or redistribution to servers or lists, or reuse of any copyrighted component of this work in other works.}
\endgroup

\begin{abstract}
Access control policies are reliability-critical configuration artifacts in cloud systems, yet administrators frequently struggle to verify that a policy permits exactly what they intend. This verification gap cannot be remedied by using LLMs to synthesize policies: we find that reasoning and non-reasoning LLMs fluently explain policy behavior but cannot reason about policy semantics with reliability-grade precision, even when the specification is the LLM's own explanation. We formulate this impasse as the Verifiable Synthesis Paradox: the verification gap persists regardless of how the policy was authored. To remedy this, we introduce \tool, a neurosymbolic tool that pairs finite-state automata with an LLM-based simplification to generate precise human-readable characterizations of requests allowed by a policy. \tool uses model counting to guarantee the fidelity of the generated characterization by rejecting characterizations that fall below a user-configured threshold in favor of the formally-derived one. On 546 AWS, 100 Microsoft Azure, and 100 Google Cloud Platform policies, \tool achieves a mean similarity score of 0.93 and a 2.7$\times$ improvement over an SMT-based baseline. In a user study, \tool raised policy-change-review accuracy from 39\% to 93\% on the hardest sub-task while reducing self-reported mental demand. We  release \tool as an open-source tool.
\end{abstract}


\section{Introduction}\label{sec:intro}
Within cloud systems, access control policies are reliability-critical configuration artifacts. Cloud systems behave exactly as configured, and when configuration diverges from administrator intent, consequences are operational failures that manifest as data exposure, service disruption, or unintended access. Misconfigurations in cloud computing services have exposed data of millions of customer accounts across Dow Jones~\cite{djleak}, Verizon~\cite{vleak}, and Microsoft Azure Cosmos DB~\cite{azureflaw}. In these incidents and others~\cite{aleak}, the underlying systems behaved as specified, but the specifications did not match what their administrators believed they had written. Thus, the central reliability question for access control is whether administrators can verify that the policy permits what they intended and nothing more. In this work, we address the verification gap that  exists regardless of how the policy was authored.


Large Language Models (LLMs) have seen great success in code generation and summarization~\cite{nam2024usingllmhelpcode,hou2024largelanguagemodelssoftware}. 
A natural extension is applying LLMs to the generation of access control policies from administrator intent. 
This raises the question of whether LLMs are capable of the precise, semantic reasoning that access control reliability demands. 
Since reliable synthesis requires reliable comprehension, in Section~\ref{sec:llm_eval} we evaluate LLM policy comprehension directly and find that LLMs produce fluent explanations of policies but fail to reason about them with reliability-grade precision. 
This gap persists even when the specification is the LLM's own explanation of the target policy. 
We formulate this gap as the \textbf{Verifiable Synthesis Paradox}: if an administrator provides a complete, unambiguous specification, an LLM is unnecessary as existing tools can compile it directly; if the specification is ambiguous, then there is no ground truth against which to verify the LLM's interpretation.

This paradox means that LLMs cannot be reliably used to synthesize correct policies. Manually written policies suffer the same consequence but for a different reason: administrators struggle to determine if a complex policy they wrote permits exactly what they intended the policy to permit.
Regardless of how the policy was authored, reliability depends on if the administrator can verify what the policy permits, in terms of what they can read.
What is needed is a tool which characterizes what a policy actually allows, in terms of what an administrator can read, so they can verify that the policy works as intended.
Existing tools can determine if a request is allowed, but they cannot characterize the set of requests the policy allows.

\begin{figure*}
\includegraphics[width=\textwidth]{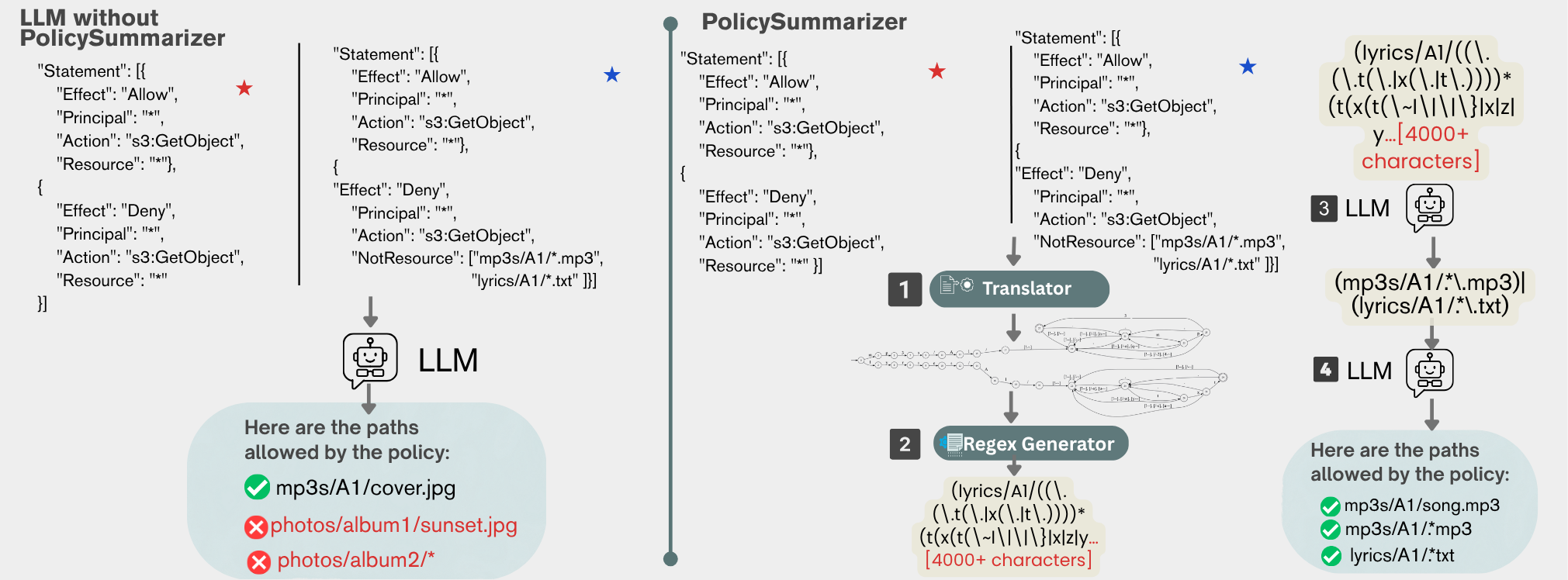}    
\caption{Overview of PolicySummarizer compared to direct LLM policy analysis. An AWS user takes an initial policy ($\textcolor{red}{\star}$) and modifies it to create a new policy ($\textcolor{blue}{\star}$), then wants to verify that the modified policy correctly allows only the intended requests. On the left, an LLM is given a policy and asked to characterize the allowed requests (Step \textbf{A}), producing an imprecise characterization (Step  \textbf{B}). On the right, PolicySummarizer encodes the policy into a finite-state automaton (Step \textbf{1}), extracts a regular expression characterizing the allowed requests (Step \textbf{2}), uses an LLM to simplify the regular expression into an interpretable one (Step \textbf{3}), producing a human-readable characterization of allowed requests (Step \textbf{4}).}
    \label{fig:tool_overview}
\end{figure*}

We address this gap with \tool (Figure~\ref{fig:tool_overview}), a tool that generates human-readable characterizations of the requests a policy allows. 
\tool uses a neurosymbolic architecture where an LLM is paired with symbolic methods to generate precise characterizations of allowed requests. 
Given a policy, \tool constructs a finite-state automaton characterizing the requests allowed by a policy, extracts a regular expression from the automaton, and uses an LLM to simplify the regular expression into a more interpretable form.
The simplified regular expression is verified against the formally derived regular expression using model counting to quantify the similarity between the two expressions. 
If the simplified regular expression falls below a user-configured similarity threshold, \tool returns the precise automata-derived regular expression of allowed requests. Beyond single-policy characterization, \tool can characterize the difference between two policies, which existing policy verifiers are not capable of. 
The neurosymbolic architecture guarantees this fidelity by construction: the LLM makes \tool's output more readable, and any simplification falling below the user-configured threshold is rejected in favor of the formally-derived regular expression.

Our contributions are as follows:
\begin{itemize}
    \item We conduct the first rigorous study of LLM comprehension and reconstruction for real-world cloud access-control policies across multiple model families. We find that LLMs produce highly consistent policy explanations, but this does not reliably translate into precise policy reasoning: request-level comprehension remains limited, and semantic reconstruction varies substantially across models.
    \item \tool, a neurosymbolic architecture for reliable access control policy analysis. \tool pairs finite-state automata and model counting with LLM-based simplification to produce a human-readable characterization of allowed requests. The architecture guarantees the fidelity of the resulting characterization by rejecting any expression which falls below the user-defined threshold in favor of the formally-derived expression.
    \item An empirical evaluation at scale of \tool across 546 AWS, 100 Microsoft Azure, and 100 Google Cloud Platform policies, achieving an average similarity score of 0.93 and a 2.7$\times$ improvement than the existing baseline. \tool also characterizes the semantic differences between policies, with mean similarity of 0.97 and 0.89 in the two directions of comparison.
    \item A user study showing \tool raises policy-change-review accuracy from 39\% to 93\% on the hardest sub-task (p < 0.001), reduces self-reported mental demand, and is preferred by 88\% of participants over reading raw JSON.
\end{itemize}
\section{Background and Preliminaries}\label{sec:overview}
In this section, we provide an overview of access control policies for cloud environments and describe the policy model that we use to analyze access control policies.

\subsection{Access Control Policies for the Cloud}
Modern cloud computing platforms allow customers to secure their data through \textit{access control policies}. 
In this work, we consider the Amazon Web Services (AWS) policy language (in Section~\ref{sec:experiments} we show our approach on other policy languages).
Formally, an AWS policy allows or denies access through declarative statements, and each policy contains one or more statements. A \textit{statement} consists of a 5-tuple (\textit{Principal, Effect, Action, Resource, Condition}) where
\begin{itemize}
    \item \textit{Principal} specifies a list of users, entities, or services
    \item \textit{Effect} $\in$ \{Allow, Deny\} specifies whether the statement allows or denies access
    \item \textit{Action} specifies a list of actions
    \item \textit{Resource} specifies a list of resources
    \item \textit{Condition} is an optional list of conditions that constrain how access is allowed or denied.
\end{itemize}
Each condition consists of a condition operator, condition key, and condition value on elements of the request context. Additionally, the AWS policy language allows the use of two special characters within strings: `*' denotes a wildcard, representing any string, and `?' which represents any possible character. For full details of the language, we direct the reader to~\cite{BBC18}.
Initially, all requests are implicitly denied. Given a request and associated policy, access is granted \textit{if and only if} there exists a statement in the policy allowing access and no statements in the policy explicitly deny access. 


\subsection{Policy Model}\label{sec:policy_model}
We use the same policy modeling as the authors in~\cite{ESL22} to encode policies into constraint formulas and briefly discuss the high level details. 
An access request is defined as a tuple $(\delta, a, r, e) \in \Delta \times A \times R \times E$, where $\Delta$ is the set of all possible principals making a request, $R$ is the set of all possible resources which access is allowed or denied, $A$ is the set of all possible actions, and $E$ is the environment attributes involved in an access request. An access control policy $\mathbb{P} = \{\rho_0, \rho_1, ... \rho_n\}$ consists of a set of rules $\rho_i$ where each rule is defined as a partial function $\rho: \Delta \times A \times R \times E \hookrightarrow \{Allow,Deny\}$. 



Given a policy $\mathbb{P} = \{\rho_0, \rho_1, ... \rho_n\}$, a request $(\delta,a,r,e)$ is granted access if and only if
\begin{small}
\begin{equation*}
    \exists \rho_i \in \mathbb{P} : \rho_i(\delta,a,r,e) = Allow \land \nexists \rho_j \in \mathbb{P} : \rho_j(\delta,a,r,e) = Deny
\end{equation*}
\end{small}
A given access request is granted access by the policy if and only if there exist a rule within the policy which allows the request and there does not exist any rule within the policy which denies the request. If a request is allowed by one rule and is denied by another rule, the request is ultimately denied.
The set of allow rules and deny rules for
$\mathbb{P}$ are defined as: 
{\small
\begin{gather*}
    \Pallow = \{ \rho_i \in \P : (\delta_i,a_i,r_i,e_i) \in \rho_i \land \rho_i(\delta_i,a_i,r_i,e_i) = \mathit{Allow}\} \\
   \Pdeny = \{ \rho_j \in \P : (\delta_j,a_j,r_j,e_j) \in \rho_j \land \rho_j(\delta_j,a_j,r_j,e_j) = \mathit{Deny}\}
\end{gather*}
}\
The requests allowed by a given policy $\P$ are the requests for which a policy rule grants access through an \textit{Allow} effect while not being denied access through a \textit{Deny} effect:
\begin{small}
\begin{equation}
\begin{split}
    \textsc{Allow}&(\P) = \{ (\delta,a,r,e) \in \Delta \times A \times R \times E : \exists \rho_i \in \mathbb{P} : \\
    &(\delta,a,r,e) \in \rho_i \land \rho_i(\delta,a,r,e) = \mathit{Allow} \\
    &\land \nexists \rho_j \in \mathbb{P} : (\delta,a,r,e) \in \rho_j \land  \rho_j(\delta,a,r,e) = \mathit{Deny} \}
\end{split}
\end{equation}
\end{small}
%



\subsection{Encoding Policies into Constraint Formulas}
The SMT encoding $\llbracket \mathbb{P} \rrbracket$ is satisfied exactly by the requests allowed by $\mathbb{P}$. Each rule $\rho$ is encoded as a conjunction of disjunctions over its principal, action, resource, and environment components, $\llbracket \rho \rrbracket = \bigwedge\nolimits_{x \in \{\delta,a,r,e\}} \bigvee\nolimits_{v \in \rho(x)} x_{smt} = v$, and the policy encoding combines allow and deny rules as $\llbracket \mathbb{P} \rrbracket = \big(\bigvee\nolimits_{\rho \in \Pallow} \llbracket \rho \rrbracket\big) \wedge \neg \big(\bigvee\nolimits_{\rho \in \Pdeny} \llbracket \rho \rrbracket\big)$. Satisfying assignments of $\llbracket \mathbb{P} \rrbracket$ correspond to requests in $\textsc{Allow}(\mathbb{P})$.


Policy rules are encoded as values for sets of $(\delta,a,r,e)$, where each set potentially grants or revokes permissions. Satisfying solutions to $\llbracket \mathbb{P} \rrbracket$ correspond to requests allowed by the policy: $ \AllowP = \{ (\delta,a,r,e) : (\delta,a,r,e) \models \smtP \}$

Semantic equivalence between policies $\Pone$ and $\Ptwo$ is decided by checking whether $F_1 = \smtPone \wedge \neg \smtPtwo$ and $F_2 = \neg \smtPone \wedge \smtPtwo$ are both unsatisfiable; when either is satisfiable, its models correspond to requests allowed by one policy not the other, which \tool later uses to characterize policy differences~\cite{ESL22}. Strict permissiveness and incomparability follow form the remaining combinations of satisfiability.

\section{Understanding the Capabilities of Large Language Models for Access Control Policies}\label{sec:llm_eval}

We evaluate LLM policy reasoning through a round-trip experiment: given a policy, we ask an LLM to explain it in natural language, and then, in a fresh context, ask an LLM to reconstruct a policy from that explanation alone. If the LLM genuinely comprehends policy semantics, the reconstruction should be semantically equivalent to the original. The experiment addresses two research questions:

\noindent\textbf{RQ1: Do LLMs comprehend policy semantics with the precision reliability-critical analysis demands?} \\
\noindent\textbf{RQ2: If we eliminate natural-language ambiguity by giving the LLM its own complete explanation as the specification, can it faithfully reproduce the policy?}

We use the Quacky policy dataset~\cite{ESL22,Quacky23}, consisting of 587 AWS access control policies.

\subsection{Experiment Workflow}
For each (policy $\P$, LLM $M$) pair, we run a 6-step workflow:
\begin{enumerate}
    \item \textbf{Original Explanation Generation ($\P \rightarrow E_{\text{orig}}$):} $M$ generates a natural language explanation $E_{\text{orig}}$ of $\P$.
    \item \textbf{Policy Reconstruction ($E_{\text{orig}} \rightarrow \P'$):} $M$ generates a new policy $\P'$ from $E_{\text{orig}}$ alone.
    \item \textbf{Reconstructed Explanation ($\P' \rightarrow E'_{\text{recon}}$):} $M$ generates an explanation $E'_{\text{recon}}$ of $\P'$.
    \item \textbf{Semantic Equivalence Check ($\P$ vs.\ $\P'$):} Encode both policies as constraint formulas and check equivalence via automata construction.
    \item \textbf{Explanation Consistency Analysis:} Compare $E_{\text{orig}}$ and $E'_{\text{recon}}$ via sentence-transformer embeddings (all-MiniLM-L6-v2) with cosine similarity.
    \item \textbf{Request Outcome Prediction ($\P \rightarrow$ Predictions):} $M$ predicts Allow/Deny for 20 test requests against $\P$ (10 expected allows, 10 expected denies), with ground truth generated systematically from $\P$.
\end{enumerate}

\begin{figure}[t]
    \includegraphics[width=\linewidth]{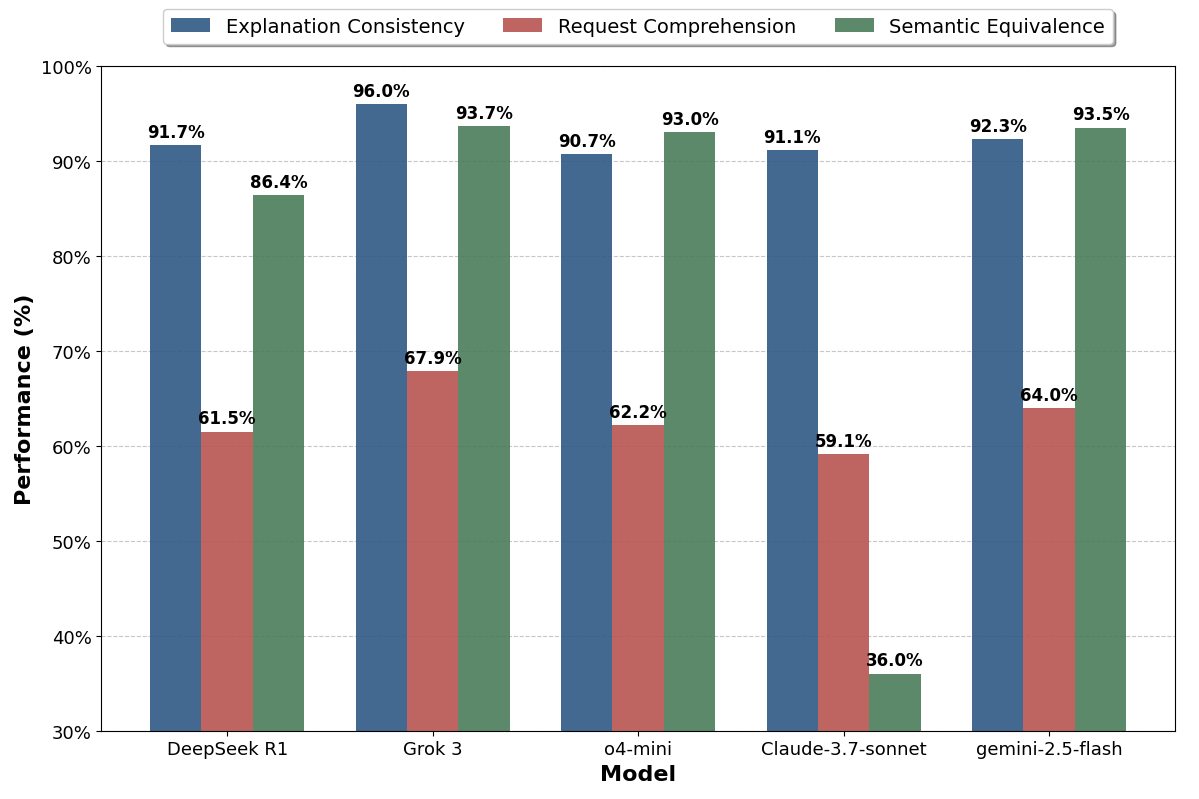}
    \caption{Comprehensive Policy Comprehension Assessment.}
    \label{fig:3_metrics}
\end{figure}

We assess LLM understanding with three metrics:
\begin{itemize}
    \item \textbf{Request Prediction:} Percentage of 20 test requests correctly labeled.
    \item \textbf{Semantic Equivalence:} Binary measure of whether $\P'$ is semantically equivalent to $\P$ (via SMT).
    \item \textbf{Explanation Consistency:} Cosine similarity between $E_{\text{orig}}$ and $E'_{\text{recon}}$ over sentence-transformer embeddings.
\end{itemize}

\subsection{Results and Discussion}
Figure~\ref{fig:3_metrics} summarizes reasoning-model results. Grok 3 led across all metrics (96.0\% explanation consistency, 67.9\% request comprehension, 93.7\% semantic equivalence). Reasoning models substantially outperformed non-reasoning counterparts on semantic equivalence (86.4--93.7\% vs.\ 15.4--45.8\%). Explanation consistency clustered tightly across reasoning models (90.7--96.0\%), while semantic equivalence varied widely (36.0--93.7\%) and request comprehension moderately (59.1--67.9\%). Claude was an outlier, pairing strong explanation consistency with weak semantic equivalence, suggesting that, for this model on this benchmark, fluent description and precise reconstruction are decoupled.

\paragraph{Answering RQ1} Reasoning models scored well on explanation consistency and semantic equivalence but underperformed on request prediction (59.1--67.9\%). \textbf{LLMs can describe policies but struggle to reason about request-level access decisions with the precision security-critical applications require.}

\paragraph{Answering RQ2} Even when given the LLM's own explanation as the specification, reconstructed policies $\P'$ were frequently not semantically equivalent to the originals --- for reasoning and non-reasoning models alike. \textbf{LLMs produce syntactically valid policies from precise specifications but introduce subtle semantic deviations.}

\paragraph{The Informal-to-Formal Asymmetry} The gap between strong explanation fidelity and weak reconstruction accuracy is consistent with an asymmetry in task difficulty: generating informal descriptions of formal artifacts is easier than synthesizing formal artifacts from informal descriptions. The former can succeed through pattern-matching over structures seen in training; the latter requires committing to specific syntactic choices whose semantic consequences the model may not fully track. This interpretation is reinforced by our request-prediction results (59.1--67.9\%), which show that even when the formal artifact is given directly, LLMs cannot reliably reason about its extensional behavior. 

\paragraph{Comparison to Existing Studies} Prior work~\cite{vatsa2025synthesizingaccesscontrolpolicies} studies LLM policy synthesis from external specifications using a single model on 47 policies, finding that structured prompts improve correctness. Our round-trip evaluation expands this to 587 policies across multiple model families and examines a complementary question: whether LLMs can faithfully reconstruct a policy when the specification is the LLM's \emph{own} explanation. Even under this favorable condition, where natural-language ambiguity is eliminated by construction, reconstruction frequently fails, indicating that comprehension limitations persist independently of the input specification's quality. Other studies~\cite{sokaccesscontrol25,plainenglishtoxacmlpolicies25,translatingnlspecintopolicyllm2024} focus on optimizing synthesis from external specifications, and~\cite{pairinghumanandaillms2024} pairs LLMs with formal specifications for enforcement. Ours is the first to systematically evaluate LLM policy comprehension as distinct from synthesis across model families.
\section{Analyzing Access Control Policies}\label{sec:analyzing_policies}
The gap identified in Section~\ref{sec:llm_eval} indicates that LLM-synthesized policies suffer from subtle correctness issues, even when prompted with the most precise explanation of the access control specification to be implemented. However, such problems are not specific to LLM-synthesized policies. Indeed, manually crafted policies can be difficult to analyze due to the complex policy logic they encode. Even simple policies are difficult to ensure that they allow only the requests intended when crafting the policy. We now detail the difficulties in analyzing both manually crafted and LLM-synthesized policies and how existing approaches are not sufficient.

\paragraph{Manually Crafted Policies} The complexity of verifying even manually crafted policies is significant. Consider the scenario depicted in Figure~\ref{fig:motivation}. A musician wants to make specific songs publicly available from an S3 bucket. They start with a policy that denies all access (left) and modify it to allow access only to resources starting with ``mp3s/A1/'' and ending with ``.mp3'' or starting with ``lyrics/A1/'' and ending with ``.txt'' (right), using a \textsc{NotResource} element in a \textsc{Deny} statement. 
Verifying that this modified policy behaves exactly as intended—allowing access \textbf{only} to the specified paths—is non-trivial. 
While AWS offers tools for testing specific requests, automatically determining and summarizing the set of allowed requests for complex policies remains a challenge. This scenario highlights the need for capabilities that can generate precise semantic summaries of policies—for instance, understanding that the policy on the right in Figure~\ref{fig:motivation} should effectively allow resources matching the regular expression \textbf{(mp3s/A1/.*\.mp3)|(lyrics/A1/.*\.txt)}. 

\begin{figure}[t]
\noindent\begin{minipage}{.5\linewidth}
\begin{lstlisting}[language=json,frame=tlrb]{Name}
"Statement": [{
   "Effect": "Allow",
   "Principal": "*",
   "Action": "s3:GetObject",
   "Resource": "*"
},{
   "Effect": "Deny",
   "Principal": "*",
   "Action": "s3:GetObject",
   "Resource": "*"
}]

\end{lstlisting}
\end{minipage}\hfill
\begin{minipage}{.5\linewidth}
\begin{lstlisting}[language=json,frame=tlrb]{Name}
"Statement": [{
   "Effect": "Allow",
   "Principal": "*",
   "Action": "s3:GetObject",
   "Resource": "*"
},{
   "Effect": "Deny",
   "Principal": "*",
   "Action": "s3:GetObject",
   "NotResource": [
      "mp3s/A1/*.mp3",
      "lyrics/A1/*.txt" ]}]
\end{lstlisting}
\end{minipage}
\caption{Initial policy allowing nothing (left), modified policy that allows access to specific resources (right). The wildcard character `$*$' represents any possible string.}\label{fig:motivation}
\end{figure}

\paragraph{Synthesized Policies}  Policies synthesized by LLMs may be \textit{mostly correct} but contain subtle semantic deviations.
We characterize two such difficulties:
\\
\noindent\textbf{Semantic Drift:} LLMs can generate policies that seem semantically similar to the intended specification but contain subtle deviations. For example, given a requirement to ``allow developers read access to project files,'' an LLM might generate a policy allowing access to ``project-*'' resources instead of ``project-data/*''. \\
\noindent\textbf{Over-Generalization from Examples:} When provided with specific examples, LLMs can generalize beyond the intended scope. A specification mentioning ``S3 buckets for logs and backups'' might result in a policy allowing access to any bucket containing ``log'' or ``backup'' in the name, rather than the specific intended buckets. This pattern leads to broader permissions than specified.

\paragraph{Limitations of Existing Approaches}
Existing tools 
such as Zelkova~\cite{BBC18}, Margrave~\cite{FKM05}, and Quacky~\cite{ESL22,Quacky23} can verify specific security properties, identify misconfigurations, or quantify permissiveness. 
However, these tools are unable to provide analysis for remedying the above difficulties, as these errors or misconfigurations tend to be more subtle, exhibiting ranges of correctness that traditional binary verification tools cannot adequately assess.
Given these unique difficulties, what is needed is a verification approach which can:
\begin{enumerate}
    \item \textbf{Characterize Requests:} Provide human-readable summary of the allowed or denied requests by a policy.
    \item \textbf{Characterize Semantic Differences:} Provide human-readable summaries of \textit{how} two policies differ in terms of the requests they allow or deny.
\end{enumerate}
This opens an ideal avenue for LLMs to be leveraged for policy analysis. As we show Section~\ref{sec:llm_eval}, LLMs are excellent at summarizing high level details of policy behaviors, but struggle to navigate the complex rules implemented by the policy. However, semantic-based approaches 
encode complex rules into mathematical structures. The mathematical structures are difficult to interpret, but this is a perfect use case where LLMs could be leveraged to interpret the mathematical structure. All that's needed is formal methods or mathematical techniques to ensure the LLM's interpretation of the mathematical structure is consistent with the semantics encoded within the structure.

\section{Semantic-Based Request Summarization using LLMs}\label{sec:framework}

\rev{Motivated by the finding that LLMs excel at interpretation but struggle with precision, we introduce \tool, a novel neurosymbolic request characterization approach which pairs LLMs with finite-state automata and model counting to generate verified characterizations of allowed requests. Specifically, we encode the set of requests allowed by a policy as strings accepted by the finite-state automata, enabling us to extract a regular expression from the automata summarizing the requests allowed by the policy. }
As regular expressions extracted from automata are often complex and uninterpretable, we leverage large language models (LLMs) to create a more simplified and interpretable regular expression. \textbf{ Crucially, we employ model counting techniques to verify the correctness of the resulting simplified regular expression.} In the case that the simplified regular expression is not sufficiently precise (below a user-configured similarity threshold) the precise, extracted regular expression is returned.

\begin{figure}[t]
    \includegraphics[width=\linewidth]{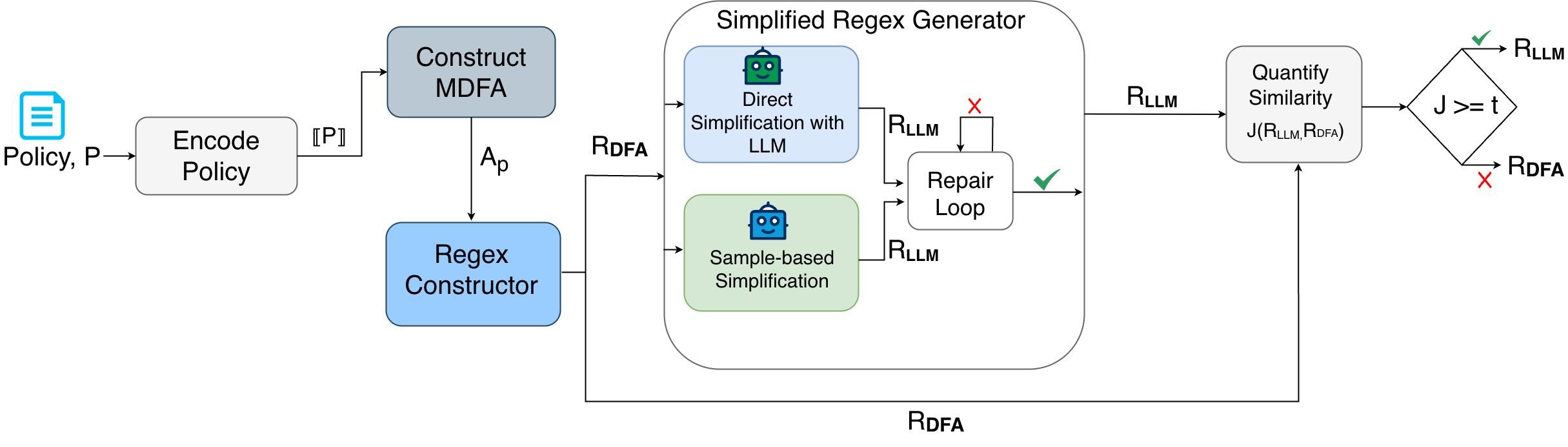}
    \caption{Overview of \tool, which encodes a policy $P$ into 
    a DFA, generates a reference regex $R_{\text{DFA}}$, and applies LLM-based 
    simplification via direct or sample-based approaches. The simplified regex $R_{\text{LLM}}$ 
    undergoes iterative repair and is accepted only if its Jaccard similarity with $R_{\text{DFA}}$ 
    exceeds threshold $t$.}
    \label{fig:policysummarizer}
\end{figure}

The workflow of \tool is shown in Figure~\ref{fig:policysummarizer}, and the summarization technique is outlined in Algorithm~\ref{alg:gensummary}.
Given a policy $\P$, the SMT formula $\smtP$ is constructed where satisfying solutions to $\smtP$ correspond to requests allowed by $\P$ (line 1). Then, a deterministic multi-track finite-state automata $\A_\P$ is constructed for $\smtP$ which characterizes the set of satisfying solutions to $\smtP$ as accepting strings in $\A_\P$ (line 2). The regular expression $\R_{\text{DFA}}$ is then extracted from $\A_\P$ (line 5), and a simplified regular expression $\R_{\text{LLM}}$ is generated through using an LLM (line 6). Model counting techniques are used to quantify the similarity between $\R_{\text{DFA}}$ and $\R_{\text{LLM}}$ (line 7). If the computed similarity is above a user-defined similarity threshold, then the simplified regular expression $\R_{\text{LLM}}$ is returned. Otherwise, the extracted complex regular expression $\R_{\text{DFA}}$ is returned. 


{\color{black}
\begin{algorithm}[t]
\footnotesize
\caption{\textsc{GenerateSummarization}($\P,b,t,n$)}\label{alg:gensummary}
\begin{flushleft}
    \textbf{Input:} Policy $\P$, model bound $b$, similarity threshold $t$, sampling size $n$ \\
    \textbf{Output:} Regular expression summarizing resources allowed by $\P$
\end{flushleft}
\begin{algorithmic}[1]
\State $\smtP = \textsc{EncodePolicy}(\P)$ 
\State $\A_\P = \textsc{ConstructMDFA}(\smtP)$ 
\State \textbf{if} $L(\A) = \emptyset$ \textbf{then} \Return $\emptyset$ 
\State \textbf{end if}
\State $\R_{\text{DFA}} = \textsc{MDFA2Regex}(\A_\P)$ 
\State $\R_{\text{LLM}} = \textsc{GenerateSimplifiedRegex}(\R_{\text{DFA}},n)$
\State $J(\R_{\text{DFA}},\R_{\text{LLM}}) = \textsc{QuantifySimilarity}(\R_{\text{DFA}},\R_{\text{LLM}},b)$
\If{$J(\R_{\text{DFA}},\R_{\text{LLM}}) \geq t$} \Return $\R_{\text{LLM}}$ 
\State \hspace{-\algorithmicindent}\textbf{else} \textbf{return} $\R_{\text{DFA}}$ 
\EndIf
\end{algorithmic}
\end{algorithm}}

\subsection{Policy Encoding and Automata Construction}
Given a policy $\P$, we first encode $\P$ into an Satisfiable Modulo Theories (SMT) formula $\smtP$ using the approach outlined in Section~\ref{sec:policy_model}, where each satisfying solution to $\smtP$ corresponds to a unique request $(\delta,a,r,e)$ allowed by $\P$. Depending on the policy language, the constraints in $\smtP$ can be boolean, linear integer arithmetic, regular expression, and string equality and dis-equality constraints. Since the expressiveness of these constraints within the context of policy analysis can be solved precisely using deterministic finite-state automata~\cite{ESL22}, we use the Automata-Based Model Counter (ABC) tool~\cite{AEB18} to construct a Multi-track Deterministic Finite-state Automaton (MDFA) $\A_\P$ which encodes the set of satisfying solutions to $\smtP$ as accepting strings in $\A_\P$. If policy $\P$ does not allow any requests, then $L(\A)=\emptyset$ and the approach stops here. 
At this point, $\A_\P$, precisely captures the set of solutions to $\smtP$, and thus the set of requests allowed by $\P$. That is:
$$ L(\A_\P) = \AllowP = \{ (\delta,a,r,e) : (\delta,a,r,e) \models \smtP \} $$

\subsection{Regular Expression Extraction}
For a policy $\P$, the satisfying strings of the constructed MDFA $\A_\P$ characterize the set of requests allowed by $\P$. While finite-state automata are useful in capturing the semantics of the policy, they do not provide a compact, straightforward understanding of the set of allowed requests. In cases where the set of requests allowed by the policy is small and finite in size, one can simply enumerate the set of requests by traversing the paths within the automaton. Aside from the simplest of policies, most policies have a rich and diverse set of requests which can be theoretically infinite (as strings can be infinitely long). We instead leverage the connection between automata and regular expressions and extract a regular expression to generate a representation of the allowed requests. 

From $\A_\P$ we use the well-known state elimination algorithm~\cite{HU79} to obtain a regular expression $\R_{\text{DFA}}$ characterizing the set of resources allowed by $\P$ (Algorithm~\ref{alg:gensummary} line 6). Since $\A_\P$ is an MDFA (which accepts tuples of strings) we first convert it to a DFA by projecting away all elements other than the resource. The state elimination method works by iteratively removing states one-by-one until only the initial and final states remain. At that point, the regular expression for the single transition remaining between the two states is returned.
This algorithm produces arbitrarily complex regular expressions which are difficult to analyze. Thus, while the extracted regular expression $\R_{\text{DFA}}$ precisely characterizes the set of resources allowed by $\P$, it is not useful for understanding the allowed resources. The extracted regular expression $\R_{\text{DFA}}$ is then fed to the simplification step.

\subsection{Regular Expression Simplification using LLMs}\label{sec:regex_simplification}
Once the extracted regular expression $\R_{\text{DFA}}$ has been generated, the next step is to transform the $\R_{\text{DFA}}$ into a simplified, more interpretable one. To do so, we leverage an LLM to assist in generating a simplified regular expression $\R_{\text{LLM}}$ which ideally retains the core semantics of $\R_{\text{DFA}}$. 
\rev{Since this process can introduce imprecision through the stochastic nature of LLMs, we provide a similarity measure using model counting techniques for quantifying how close $\R_{\text{LLM}}$ is to the precise extracted regular expression $\R_{\text{DFA}}$ (discussed below). We describe two approaches for regex simplification: Direct Simplification, and Sampling-Based Simplification. After each LLM call we check the regular expression for syntactic validity. If it is invalid, we re-prompt the LLM with the original prompt and the error message appended. This process is repeated up to $k$ times ($k=5$ in our approach). If it is still invalid after $k$ attempts, $\R_{\text{DFA}}$ is returned.}

\rev{\paragraph{Direct Simplification} In the direct simplification approach, we feed the LLM the extracted regular expression in its entirety and prompt it to generate a simplified, semantically equivalent regular expression. This approach avoids the parameter-sensitivity in the sampling-based simplification approach. Current generation models are more suited to this task, and in Section~\ref{sec:experiments} we show that these models can perform direct simplification and rarely trigger the repair loop, and produce more semantically accurate regular expressions.}

\rev{\paragraph{Sampling-Based Simplification}
In the sampling approach we randomly sample a set of accepting strings from $\A_\P$ and query an LLM to generalize the sample of strings into a simplified regex.}
As sampling from an DFA requires dealing with cycles (and thus infinite loops), we opt instead to randomly sample accepting strings from the extracted regular expression $\R_{\text{DFA}}$. We sample strings via recursive descent over the regex AST, choosing union branches uniformly and using a geometric-decay terminiation for Kleene-star expansions to ensure finite samples.
{\color{black}
\begin{algorithm}[t]
\footnotesize
\caption{\textsc{QuantifySimilarity}($\R_{\text{DFA}},\R_{\text{LLM}},b$)}\label{alg:genregexllm}
\begin{flushleft}
    \textbf{Input:} Regular expressions $\R_{\text{DFA}}, \R_{\text{LLM}}$, model counting bound $b$ \\
    \textbf{Output:} Similarity measure $J(\R_{\text{DFA}},\R_{\text{LLM}})$ quantifying similarity between $\R_{\text{DFA}}, \R_{\text{LLM}}$
\end{flushleft}
\begin{algorithmic}[1]
\State $\A_{\R_{\text{DFA}}} = \textsc{ConstructDFA}(\R_{\text{DFA}})$
\State $\A_{\R_{\text{LLM}}} = \textsc{ConstructDFA}(\R_{\text{LLM}})$
\State $\A_n = \textsc{ConstructDFA}(\A_{\R_{\text{DFA}}} \wedge \A_{\R_{\text{LLM}}})$ 
\State $\A_d = \textsc{ConstructDFA}(\A_{\R_{\text{DFA}}} \vee \A_{\R_{\text{LLM}}})$ 
\State $\mcount{\A_n} = \textsc{CountModels}(\A_n,b)$ 
\State $\mcount{\A_d} = \textsc{CountModels}(\A_d,b)$ 
\State \Return $\frac{\mcount{\A_n}}{\mcount{\A_d}}$ 
\end{algorithmic}
\end{algorithm}
}
\paragraph{Quantifying Similarity of Simplified Regular Expression}
We compare the precise regular expression $\R_{\text{DFA}}$ with the simplified regular expression $\R_{\text{LLM}}$ using quantitative analysis techniques. We use the Jaccard Similarity Measure (JSI), otherwise known as Intersection-Over-Union (IoU), to measure the similarity between the sets of strings represented by $\R_{\text{DFA}}$ and $\R_{\text{LLM}}$. \rev{Note that SMT-based reasoning tools (e.g., OSTRITCH~\cite{hague2025ostrich2solvercomplexstring}) can be used to perform binary equivalence of regular expressions, but our approach requires quantifying the degree of similarity rather than binary equivalence. This is because the two regular expressions could differ by one string, yet a binary equivalence would show they are not equivalent. Quantifying similarity instead provides a degree of similarity.} The process is shown in Algorithm~\ref{alg:genregexllm}.
We first convert $\R_{\text{DFA}}$ and $\R_{\text{LLM}}$ to automata and perform model counting queries to count the number of strings shared between the set of strings represented by $\R_{\text{DFA}}$ and the set of strings represented by $\R_{\text{LLM}}$. In other words, we compare their corresponding DFAs $\A_{\R_{\text{DFA}}}$ and $\A_{\R_{\text{LLM}}}$.
Specifically, we measure the similarity of the two sets $L(\A_{\R_{\text{DFA}}})$ and $L(\A_{\R_{\text{LLM}}})$ by 
\begin{equation}
    J(\R_{\text{DFA}},\R_{\text{LLM}})=\frac{\mcount{L(\A_{\R_{\text{DFA}}} \land \A_{\R_{\text{LLM}}})}}{\mcount{L(\A_{\R_{\text{DFA}}} \vee \A_{\R_{\text{LLM}}})}}
\end{equation}
We use model counting techniques to compute $\mcount{L(\A_{\R_{\text{DFA}}} \land \A_{\R_{\text{LLM}}})}$ and $\mcount{L(\A_{\R_{\text{DFA}}} \vee \A_{\R_{\text{LLM}}})}$ by counting accepting paths in the corresponding DFAs, again bounding the lengths of accepting paths to some bound $b$ (otherwise the sets would be infinite). Note that $0 \leq J(\R_{\text{DFA}},\R_{\text{LLM}}) \leq 1$: higher values correspond to more similar regular expressions, while lower values of correspond to less similar regular expressions. 

For the modified policy in Figure~\ref{fig:motivation}, \tool computes an extracted regular expression from the DFA which contains over 4000 characters. Using direct simplification, \tool generates the simplified regular expression \textbf{(mp3s/A1/.*\.mp3)|(lyrics/A1/.*\.txt)}, with our quantitative similarity analysis giving a similarity score of $1.0$. 

\paragraph{Summarizing Differences Between Policies}\label{sec:comparing_policies}
Our resource summarization technique can also be used to \textbf{characterize the semantic differences between policies}. We compute the semantic equivalence between two policies $\Pone$ and $\Ptwo$ by encoding $\Pone,\Ptwo$ into $\smtPone,\smtPtwo$ and constructing the formulas $F_1 = \smtPone \wedge \neg \smtPtwo$ and $F_2 = \neg \smtPone \wedge \smtPtwo$. We can then construct DFAs $\A_{F_1}$ and $\A_{F_2}$ where strings accepted by $\A_{F_1}$ correspond to requests allowed by $\Pone$ but not by $\Ptwo$, and strings accepted by $\A_{F_2}$ correspond to requests allowed by $\Ptwo$ but not by $\Pone$. The technique then works similarly as it does for a single policy, except in this case two sets of extracted and simplified regular expressions are generated. The extracted and simplified regular expressions from $\A_{F_1}$ summarize the requests allowed by $\Pone$ but not by $\Ptwo$, while the extracted and simplified regular expressions from $\A_{F_2}$ summarize the requests allowed by $\Ptwo$ but not by $\Pone$.

\section{Experiments}\label{sec:experiments}
We now discuss our experimental evaluation of \tool and the neurosymbolic request summarization technique within it. Unless otherwise stated, \tool use the Direct Simplification approach as the default simplification approach. We aim to answer the following questions: \\
\noindent\textbf{RQ1:} Can \tool successfully generate precise resource characterizations for policies? \\
\noindent\textbf{RQ2:} \rev{How does Direct Simplification compare with Sample-based Simplification?} \\
\noindent\textbf{RQ3:} How does \tool 's summarization compare against baseline approaches? 

\subsection{Experimental Setup}
To generate the request summarization, \tool uses \quacky~\cite{Quacky23} to translate policies into constraints, and uses the Automata-Based Model Counter (ABC)~\cite{AEB18} to construct automata and perform model counting queries. We modified ABC to extract regular expressions from automata and to compute the similarity of the generated regular expressions and extracted regular expressions. We ran all experiments on a machine with an AMD® Ryzen Threadripper Pro 5955wx processor with 16 cores and 32 threads. It has 256.0 GiB of memory and an NVIDIA Corporation GA102GL [RTX A6000] GPU. The operating system is Ubuntu 22.04.4 LTS, 64-bit.

\paragraph{Policy Dataset} \rev{We show the generalizability of \tool through benchmarking across AWS, Microsoft Azure, and Google Cloud Platform policies. For the AWS policies, we use the original 587 policies from the Quacky benchmark~\cite{ESL22,Quacky23} which consists of 41 AWS access control policies taken from user forums, extended to 587 policies through policy mutations. For Azure and GCP policies, we use 100 built-in role definitions collected from an IAM Dataset GitHub Repository~\footnote{https://github.com/iann0036/iam-dataset}, which is scraped from online sources and Cloud APIs and is actively maintained. We generate one role assignment (Azure) and role binding (GCP) per role definition, randomly pairing each with principals and resource scopes sampled from a predefined pool of representative identities and resource hierarchy paths, yielding 100 policies for Azure and 100 for GCP.}

\paragraph{Model Counting Bounds} Model counting formulas containing string constraints require a finite bound on the length of the string due to strings of infinite length. \rev{In our experiments, we limit the maximum length of strings to be 100 for AWS policies and 250 for Azure/GCP policies}. Additionally, we use a subset of the ASCII encoding to represent strings: rather than each character having 256 possibilities, and we limit characters to take on only printable values within the ASCII encoding. 

\paragraph{Large Language Models} 
For generating the simplified regular expression $\R_{\text{LLM}}$, we used Claude-4-Sonnet as we determined that it worked best for regular expression generation. In our evaluation, it consistently generated simplified regular expressions with the highest similarity scores. The selection of Claude Sonnet 4 was also principally driven by its large context window (200K), which enables the model to process a substantial number of example strings. Note that due to the statistical nature of LLMs, we query the LLM 5 times and take the best result of the 5 tries. Each LLM query generates a regular expression which must be verified using \tool.


\subsection{Results}

\begin{figure}[t]
\includegraphics[width=\linewidth]{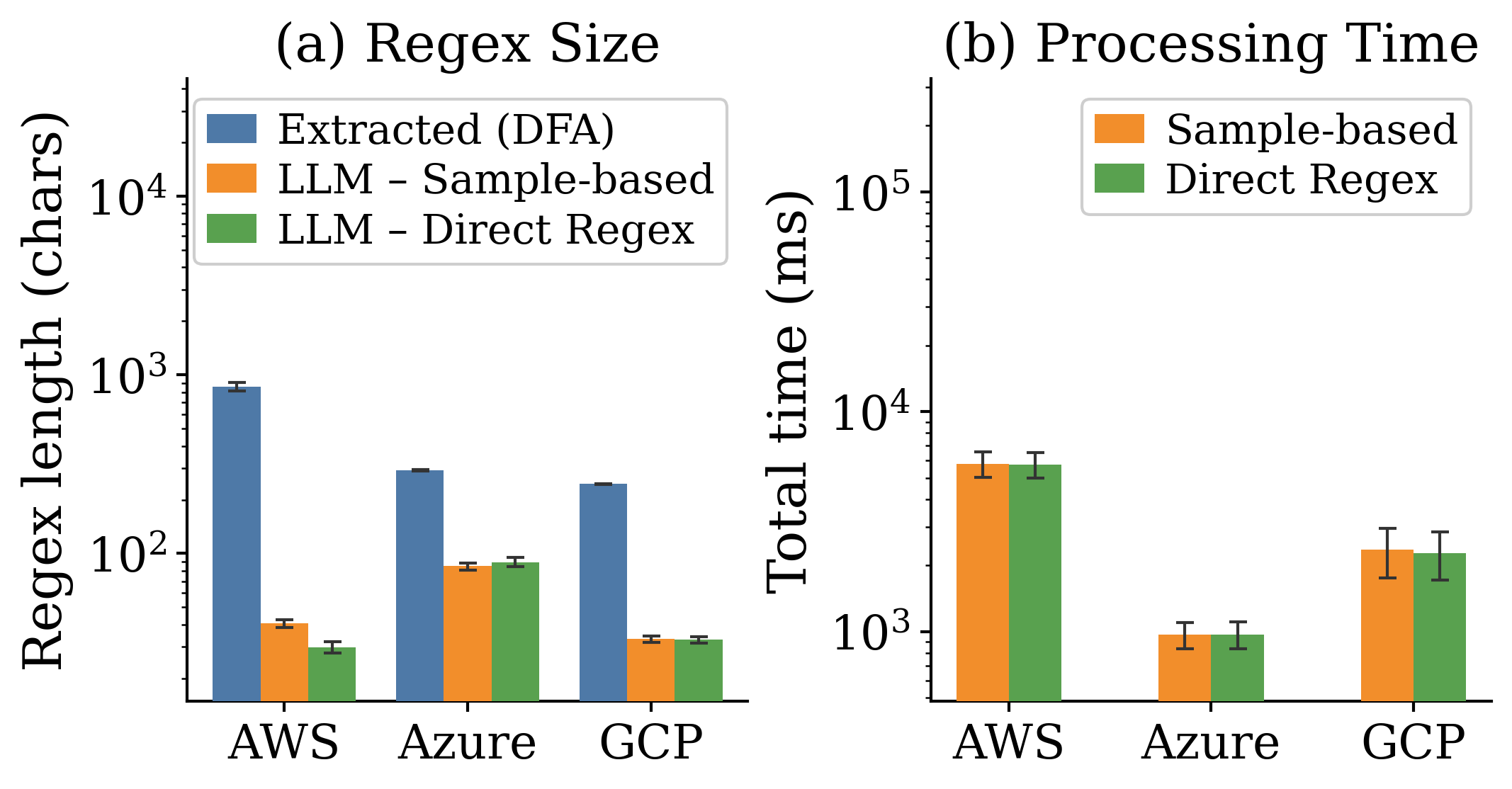}
\caption{Regex compactness and processing cost of the two summarization approaches.
\textbf{(left panel)} Mean character length of the DFA-extracted reference regex versus the
LLM-generated summary regex produced by the \textit{Sample-based} and \textit{Direct Regex}
approaches (log scale).
\textbf{(right panel)} Mean total processing time per policy (SMT solving $+$ model counting,
log scale). Total processing time per policy is ${\sim}5.8$\,s for AWS, ${\sim}2.4$\,s for GCP,
and ${\sim}1.0$\,s for Azure.
Error bars denote $\pm1$ standard error of the mean.}
\label{fig:resource_characterization_results}
\end{figure}

\textbf{RQ1:} \textit{Can \tool successfully generate precise resource characterizations for policies?} We use \tool to generate summarizations for all 587 AWS, 100 Microsoft Azure, and 100 GCP policies shown in Figure~\ref{fig:resource_characterization_results}. We observe that \tool produces simplified regular expressions significantly shorter than the original DFA-extracted expressions with the Direct Regex summarization approach \textbf{while maintaining high similarity scores with a mean value of 99.0\%, 94.9\% and 91.0\% (for a total mean of 92.6\%) for Azure, GCP and AWS, respectively}, demonstrating that the simplified regexes retain the essential core semantics of the extracted regular expressions across the three cloud providers. Table \ref{tab:jaccard_dist} shows the overall distribution of the similarity scores of LLM-generated Vs DFA-synthesized regexes across AWS, Azure, and GCP policies. Out of the 587 AWS policies, \tool found that 41 (7.0\%) policies were unsatisfiable; hence, they are excluded from the analysis. In three instances, the LLM generated regular expression was syntactically invalid; in all three cases, a single repair loop iteration fixed the syntactic issues.

Beyond single-policy summarization, we also evaluate \tool for policy difference characterization. Recall that the \quacky dataset contains 41 original AWS policies which were extended using mutations to 587 policies. From this dataset, after filtering out policies that allowed no requests, we constructed 546 policy pairs where each original policy is paired with its mutated version. For each policy pair, \tool computes the Jaccard similarity between the DFA-synthesized regular expression and the LLM generated one for both directions of the policy comparison, where the first direction ($\smtPone \land \neg \smtPtwo$) and second direction ($\neg \smtPone \land \smtPtwo$) characterize requests allowed by one policy but not the other. Of these, \tool found that 302 mutants (55.3\%) were more permissive than their originals, 48 (8.8\%) were less permissive, 44 (8.1\%) were incomparable, and 126 (23.1\%) were semantically equal; 26 pairs (4.8\%) timed out during SMT solving and were excluded. Table~\ref{tab:jaccard_dir} reports the Jaccard similarity between DFA-synthesized and LLM-generated regexes restricted to satisfiable pairs. \tool achieves mean Jaccard similarities of 0.967 and 0.894 for the two directions, respectively, with both attaining a median Jaccard of 1.0. The results show that \tool generalizes beyond single-policy summarization to the task of precisely characterizing policy differences, identifying requests one policy allows that another does not, and vice versa. 


\definecolor{shadeblue}{RGB}{235, 241, 251}
\definecolor{perfectgreen}{RGB}{200, 230, 200}
\definecolor{bestcell}{RGB}{150, 210, 150}
\begin{table}[t]
\centering

\caption{Jaccard similarity distribution by dataset and extraction method.
         \emph{Perfect} = exact match ($=1.0$); \emph{Zero} = none or negligible overlap ($=0.0$).
         Shading highlights the dominance of perfect matches.}
\label{tab:jaccard_dist}
\footnotesize
\setlength{\tabcolsep}{6pt}
\renewcommand{\arraystretch}{1.3}
\resizebox{\linewidth}{!}{
\begin{tabular}{lrrr rrrrrrr}
\toprule
Dataset & $n$ & Perfect & Zero
  & \multicolumn{7}{c}{Jaccard Similarity Distribution} \\
\cmidrule(l){5-11}
 & & &
  & {(0, .2)} & {[.2, .4)} & {[.4, .6)} & {[.6, .8)}
  & {[.8, .9)} & {[.9, .95)} & {[.95, 1)} \\
\midrule
\rowcolor{shadeblue}
AWS String   & 546 & \cellcolor{perfectgreen}392 & 50 & 19 & 6 & 8 & 24 & 0 & 1 & 46 \\
\rowcolor{shadeblue}
AWS Regex    & 546 & \cellcolor{bestcell}\textbf{497} & 27 & 14 & 1 & 1 & 2 & 1 & 0 & 3 \\
Azure String & 100 & \cellcolor{perfectgreen}93  & 4  & 3 & 0 & 0 & 0 & 0 & 0 & 0 \\
Azure Regex  &  100 & \cellcolor{bestcell}\textbf{98}  & 2  & 0 & 0 & 0 & 0 & 0 & 0 & 0 \\
\rowcolor{shadeblue}
GCP String   & 100 & \cellcolor{perfectgreen}92  & 8  & 0 & 0 & 0 & 0 & 0 & 0 & 0 \\
\rowcolor{shadeblue}
GCP Regex    &  100 & \cellcolor{bestcell}\textbf{94}  & 5  & 1 & 0 & 0 & 0 & 0 & 0 & 0 \\
\bottomrule
\end{tabular}}
\end{table}

\begin{figure}[t]
    \centering
    \includegraphics[width=\columnwidth]{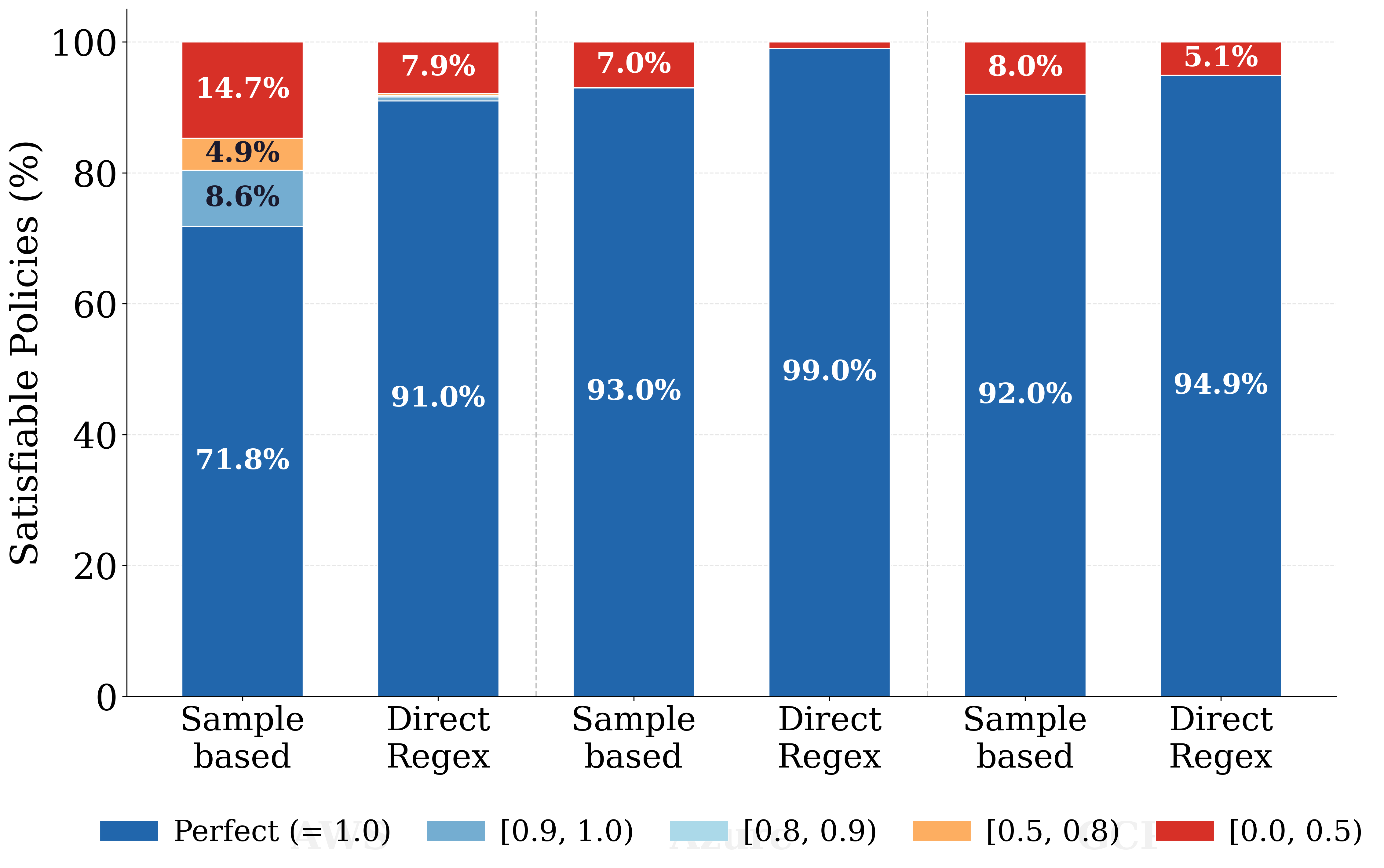}
    \caption{Distribution of Jaccard similarity scores for generated summarizations, grouped by cloud provider (AWS, Azure, GCP) and method (sample-based vs. direct regex).}
    \label{fig:similarity_results_all_policies}
\end{figure}

\rev{
\textbf{RQ2:} \textit{How does Direct Simplification compare with Sample-based Simplification?} To answer RQ2, we use \tool to evaluate two complementary approaches to regex summarization discussed in Section~\ref{sec:regex_simplification}. Figure~\ref{fig:similarity_results_all_policies} shows the results of the two approaches over AWS, Azure, and GCP policies. Direct Simplification (Direct Regex) yields higher summarization similarity with 91.0\%, 99.0\%, and 94.9\% for AWS, Azure, and GCP policies, respectively, while Sample-based simplification achieves relatively lower similarity scores of 71.8\%, 93.0\%, and 92.0\% for AWS, Azure, and GCP. The higher similarity score of the Direct Simplification approach indicates that \tool provides more accurate summarizations than the Sample-based approach.}

\begin{figure}[t]
    \includegraphics[width=\linewidth]{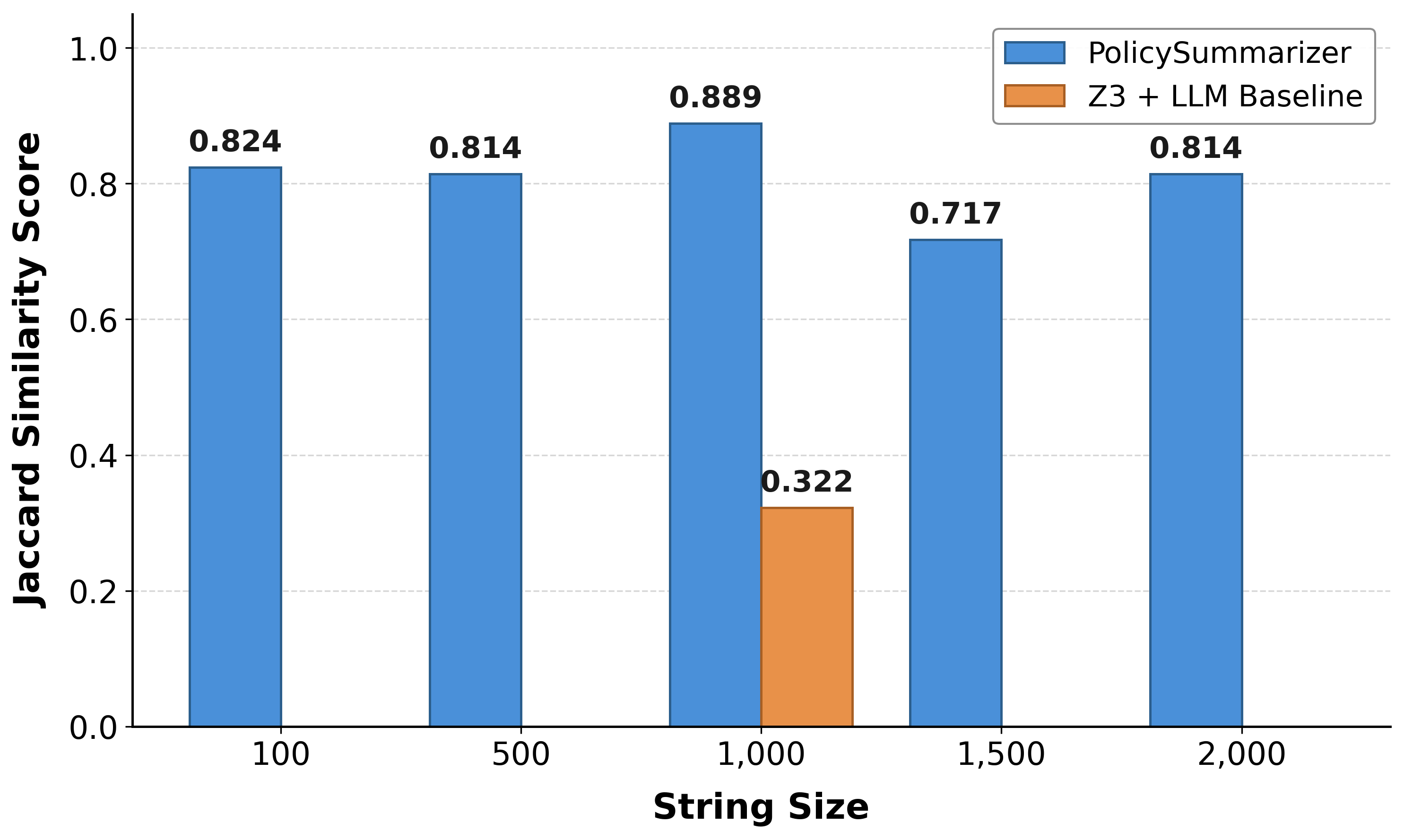}
    \caption{Comparison of average similarity measure for the Sampling-based simplification approach within \tool for varying sample sizes. For string size of 1,000, we report the Z3+LLM Baseline compared to the result given by \tool.}
    \label{fig:similarity_by_size}
\end{figure}

We further analyze the relationship between string size and summarization quality shown in Figure~\ref{fig:similarity_by_size} over the 41 original policies in the Quacky dataset. We observe that Jaccard similarity scores rise from approximately 0.824 with 100 sample strings to over 0.889 with 1000 strings and then decrease beyond that. The improvement from 100 to 1000 is also not uniform and exhibits variation at intermediate sample sizes. This shows that there is a sweet spot that leads to better regex summarization quality beyond which the llm might overfit, indicating that optimal string size selection is crucial.

\definecolor{shadeblue}{RGB}{235, 241, 251}
\definecolor{bestcell}{RGB}{150, 210, 150}
\definecolor{warncell}{RGB}{250, 210, 210}
\begin{table}[t]
\centering
\caption{Jaccard similarity between DFA-synthesized and LLM-generated regexes restricted to pairs where the given comparison is satisfiable. $n_J$ counts the number of times the LLM generated a valid regular expression; in 1 case (6 cases resp), the generated regular expression was syntactically invalid.}
\label{tab:jaccard_dir}
\footnotesize
\setlength{\tabcolsep}{6pt}
\renewcommand{\arraystretch}{1.3}
\resizebox{\linewidth}{!}{
\begin{tabular}{l rr rr rr rr}
\toprule
Comparison
  & {$n_\text{SAT}$} & {$n_J$}
  & {Mean $J$} & {Median $J$}
  & \multicolumn{2}{c}{Perfect ($J=1.0$)}
  & \multicolumn{2}{c}{Zero ($J=0.0$)} \\
\cmidrule(lr){6-7}\cmidrule(lr){8-9}
  & & & & & {$n$} & {(\%)} & {$n$} & {(\%)} \\
\midrule
\rowcolor{bestcell}
$\smtPone \land \neg \smtPtwo$ &  92 &  91 & \textbf{0.967} & \textbf{1.000} &  \textbf{88} & \textbf{96.7} &  3 &  3.3 \\
 $\neg \smtPone \land \smtPtwo$ & 346 & 340 & 0.894 & 1.000 & 298 & 87.6 & \cellcolor{warncell}25 & \cellcolor{warncell}7.4 \\
\bottomrule
\end{tabular}}
\end{table}

\textbf{RQ3:} \textit{How does \tool's summarization approach compare against baseline approaches?}
No existing policy verifier directly generates summaries of allowed requests, so we emulate a natural SMT-based baseline: for each policy $\P$, we generate $\smtP$, use Z3 to enumerate 1000 satisfying models, and prompt Claude-4-Sonnet to generalize those examples into a regular expression. Because the Z3 baseline only enumerates satisfying assignments, it supports sample-based generalization but does not produce a symbolic language representation of all satisfying resources. Direct Simplification is therefore only available in \tool, where the automata pipeline extracts $\R_{\text{DFA}}$ before invoking the LLM. While prior regex-synthesis systems can generate regexes from examples or natural-language descriptions, they solve a different problem and do not provide a drop-in replacement for extracting a complete policy-language characterization from an SMT formula.

For fairness, we compare the Z3 baseline against the sample-based variant of \tool using the same sample size ($n=1000$), rather than against \tool's stronger Direct Simplification approach. This makes the comparison conservative: as shown in RQ2, Direct Simplification is the better-performing \tool configuration, but it has no analogue in a model-enumeration baseline that lacks $\R_{\text{DFA}}$.

To isolate the effect of the input representation, the baseline and sample-based \tool use the same downstream procedure: the same LLM, prompt structure, best-of-5 selection, syntactic repair loop, and Jaccard-based evaluation. They differ only in what is given to the LLM: 1000 Z3-enumerated satisfying models for the baseline versus 1000 strings sampled from the DFA-extracted regex for \tool.

Across the 41 original policies, the sample-based variant of \tool achieved an average similarity score of 0.889, compared to 0.322 for the Z3-based baseline, a 2.7$\times$ improvement. This gap indicates that even when \tool is restricted to the same sample-generalization setting as the baseline, samples drawn from the automata-derived language representation produce substantially more accurate summaries than Z3-enumerated satisfying models. Since Direct Simplification performs better than sample-based simplification, this comparison likely understates the advantage of \tool's default configuration. Finally, computing the Jaccard similarity itself requires the automata/model-counting machinery in \tool, since the generated summary must be compared against the policy language rather than only against the sampled examples.

\section{Evaluation: User Study}
\label{sec:user_study}
We recruited 41 participants through our software engineering and academic networks, including graduate students from research labs and current and former industry software engineers to evaluate whether \tool helps users audit access-control policies more accurately and efficiently than reading raw JSON.
Participants were recruited through software engineering and academic networks, including graduate students and current or former industry software engineers. Participation was voluntary and uncompensated; the study was conducted through an anonymous web-based survey, no personally identifiable information was collected, and participants provided informed consent before beginning. Participants reported moderate experience with AWS environments (mean: 3.07/5) and AWS IAM (mean: 3.05/5).

\subsection{Methodology and Tasks}

We used a fixed-order within-subjects design. Participants first completed each task using raw JSON policies, then completed the corresponding task using \tool. We used this order because \tool explicitly reveals the effective policy scope; presenting it first would spoil the baseline condition. As a result, timing results should be interpreted conservatively, since participants saw the policies once before using the tool.

Participants acted as security auditors for two scenarios designed to exercise common policy-review failure modes. In the \textbf{Policy Audit} task, participants inspected a policy with several restrictive decoy statements and one broad wildcard statement granting access to all S3 buckets. They were asked to list all S3 resource paths allowed by the policy and provide an example allowed path outside the reports bucket. In the \textbf{Policy Change Review} task, participants compared two syntactically different but semantically equivalent policies and were asked to list paths that became accessible and paths where access was revoked. The first task tests whether users can characterize the full scope of an over-permissive policy; the second tests whether users can avoid inferring semantic changes from syntactic refactoring.

\subsection{Results}

We graded participants' answers against the mathematical ground truth of the policies. Table~\ref{tab:accuracy} reports paired accuracy under raw JSON and \tool, with McNemar exact tests for each sub-task.

\begin{table}[t]
\centering
\caption{Task Accuracy: Raw JSON vs. \tool ($N=41$). $p$-values from McNemar's exact test on paired correctness.}
\label{tab:accuracy}
\begin{tabular}{lccc}
\toprule
\textbf{Sub-Task} & \textbf{Raw JSON} & \textbf{Summarizer} & \textbf{$p$} \\
\midrule
\textbf{Audit: List Resources} & 25/41 (61.0\%) & 41/41 (100.0\%) & $<.001$ \\
\textbf{Audit: Outside Example} & 29/41 (70.7\%) & 35/41 (85.4\%) & $.180$ \\
\textbf{Change: Added Paths} & 18/41 (43.9\%) & 39/41 (95.1\%) & $<.001$ \\
\textbf{Change: Removed Paths} & 16/41 (39.0\%) & 38/41 (92.7\%) & $<.001$ \\
\bottomrule
\end{tabular}
\end{table}

\tool improved correctness across all four sub-tasks. In the Policy Audit task, many participants using raw JSON could identify that some outside access existed, but fewer could characterize the full exposure: 29/41 (70.7\%) named one leaked outside bucket, while only 25/41 (61.0\%) correctly listed the full allowed scope. With \tool, all participants correctly characterized the policy scope. In the Change Review task, raw JSON led many participants to infer changes that were not present: only 18/41 (43.9\%) correctly identified that no paths were added, and only 16/41 (39.0\%) correctly identified that no paths were removed. With \tool, these increased to 39/41 (95.1\%) and 38/41 (92.7\%), respectively (Note: Although a binary verifier could prove equivalence in this specific task, the purpose of the task is to test whether semantic summaries help users interpret policy changes; when two policies are not equivalent, \tool also reports the concrete semantic delta rather than only a binary verdict). These results highlight the value of semantic policy summaries: syntactically different policies can be difficult to compare manually even when they are semantically equivalent.

Participants also completed tasks faster with \tool, although we treat timing as secondary evidence because of the fixed-order design. Average completion time decreased from 146.8s to 108.2s for Policy Audit and from 185.8s to 105.7s for Change Review. Perceived mental demand, measured using the NASA-TLX mental demand subscale adapted to a 7-point Likert scale, decreased from 5.17 under raw JSON to 3.27 with \tool (paired Wilcoxon signed-rank test, $p<.001$). Confidence on the Change Review task increased from 2.81/5 to 4.00/5, and 36/41 (87.8\%) participants preferred \tool over raw JSON.

Free-text responses were consistent with the quantitative results. Participants emphasized that raw JSON made policy intersections and semantic diffs hard to compute manually, while \tool made the effective scope and changes explicit. One participant summarized the experience: \textit{``Task 2 was tricky without the tool. The diffs in JSON are hard to spot.''}

\section{Discussion}\label{sec:discussion}

Our results suggest a narrow but useful role for LLMs in reliable policy analysis: they should interpret formally derived artifacts, not act as the source of semantic truth. Section~\ref{sec:llm_eval} shows that models can produce consistent policy explanations while still failing to reconstruct semantically equivalent policies. For access-control artifacts, this gap is dangerous because small semantic deviations can silently create over-permissive policies.

\tool is designed around this constraint. The trusted object is the automata-derived characterization $\R_{\text{DFA}}$; the LLM-generated regex $\R_{\text{LLM}}$ is only a presentation layer. It is accepted only when model counting shows sufficient agreement with $\R_{\text{DFA}}$, and otherwise the tool returns the precise extracted regex. This also explains why direct regex simplification outperforms sample-based simplification: the LLM is more reliable when grounded in a complete symbolic representation than when asked to generalize from examples, a setting that can reproduce the same over-generalization failures seen in policy synthesis.

In practice, the similarity threshold $t$ determines how conservative the tool should be. High-assurance reviews should use $t=1.0$, requiring exact agreement under the chosen bound before trusting the simplified summary. Lower thresholds may still be useful for exploratory auditing, but should be treated as guidance rather than evidence of complete semantic equivalence. Although this paper focuses on resources, the underlying MDFA encodes full request tuples, so the same projection-based approach can be extended to principals, actions, and conditions.

\subsection{Threats to Validity}

Our evaluation has four main threats to validity. First, model counting over string constraints requires a finite length bound. We use $b=100$ for AWS and $b=250$ for Azure and GCP, so the reported similarity scores measure agreement within those bounded languages rather than unbounded regular-language equivalence. We chose these bounds to cover the resource identifiers observed in our datasets, but policies with longer or unusual resource strings may require larger bounds.

Second, our LLM results depend on the models and prompts evaluated. We include models from multiple providers and compare reasoning and non-reasoning models, but the LLM landscape changes quickly and prompt design can affect both policy reconstruction and regex simplification. The broader conclusion is therefore not tied to any single model: policy-generation claims should be checked semantically rather than inferred from explanation quality.

Third, our policy datasets may not capture all production policy structures. The AWS evaluation uses the Quacky benchmark and its mutations, while the Azure and GCP evaluations use built-in role definitions with generated assignments or bindings. These datasets exercise important policy features, but real deployments may include permission boundaries, cross-account trust, organizational conventions, and service-specific idioms that require further study.

Finally, the user study uses a fixed-order design: participants inspected raw JSON before using \tool on the same tasks. This ordering was necessary because \tool reveals the effective policy scope and would otherwise spoil the baseline condition, but it may introduce a learning effect. To bound this effect, we conservatively attribute the entire non-significant gain on the Audit: Outside Example task to repeated exposure, yielding a net self-correction rate of $\hat{\rho}=6/12=0.50$. Applying this rate to the remaining tasks gives learning-only accuracy ceilings of 33/41, 29.5/41, and 28.5/41, below the observed \tool accuracies of 41/41, 39/41, and 38/41. Matching the observed gains would require $\rho>0.88$ across all three tasks. Thus, while fixed ordering may inflate timing improvements, it is unlikely to explain the primary correctness gains. External validity is also limited by the two controlled policy scenarios and by recruitment from software engineering networks; results may differ for broader administrator populations or production policies with different structures.

\section{Related Work}\label{sec:related}

Access control~\cite{samarati01policiesmodelsmechanisms,sandhu94access,sandhu96audit}, access control policy languages~\cite{abadpeiro99plas,jajodialogical,jajodia01multiplepolicies,jajoida97unified}, and verification approaches~\cite{DBLP:conf/cade/DoughertyFK06,
DBLP:journals/sttt/HughesB08,schaad:lightweight,zao:rbac} have seen extensive research. There has also been approaches for assisting the creation of policies~\cite{DBLP:conf/sacmat/FislerK10,
Egelman:2011:OID:1978942.1979280}. Recent work has investigated machine learning and LLMs for synthesizing access control policies~\cite{vatsa2025synthesizingaccesscontrolpolicies,subramaniam2025deploiapplyingnl2sqlsynthesize,translatingnlspecintopolicyllm2024,translatingnlspecintopolicyllm2024,sokaccesscontrol25,plainenglishtoxacmlpolicies25} or enforcing access control~\cite{promptingllmenforcevalidate2024, pairinghumanandaillms2024}. Our work differs in that ours is the first to investigate LLM comprehension and synthesis in the context of access control for the cloud, and only \tool can generate request summarizations for policies.

Constraint solving approaches for policy analysis has been a subject of recent research, particularly in the proprietary Zelkova tool used within AWS~\cite{BBC18} which translates policies into SMT formulas. The work in~\cite{ESL22,Quacky23} extended this work by introducing quantitative analysis for reasoning about the permissiveness of policies, and was most recently applied to quantitative policy repair~\cite{QuackyRepair}. While our work uses the same policy model to encode policies into formulas, our work is the first to apply quantitative analysis techniques to verify policies synthesized using LLMS and generate precise request summarizations. Margrave~\cite{FKM05} is a tool that analyzes XACML policies using multi-terminal decision diagrams. Margrave goes beyond binary/ternary differential analysis, allowing a user to write general-purpose queries over changes to a policy. In a later work~\cite{NBD2010}, Margrave uses a SAT solver to enumeratively produce sets of solutions to queries.

There is work in generating a semantic representation of requests allowed by policies. In a work closely related to ours~\cite{BCC20}, Zelkova was extended to verify if policies were Trust Safe (i.e., blocks public access and does not allow untrusted requests). The authors also introduced a syntactic approach for determining the set of requests allowed by the policy. Our work provides a more precise, semantic-based approach for representing requests. Additionally, the work in~\cite{BCC20} is proprietary and not available to compare against. In addition to introducing policy repair, ~\cite{QuackyRepair} introduces a heuristic for extracting regular expressions from automata without loops for understanding policies. Our approach is capable of handling loops while generating a useful, precise regular expression. 

Automatically generating regexes has also been the subject of recent works. InfeRE~\cite{inferRE} uses LLMs to generate regexes from natural language descriptions. Our work differs from this in that there is no natural language description from which we can generate a regex from. Example-based approaches for generating regexes exist~\cite{alphaRegex,regexGenetic,flashRegex,inferConciseRE}, though our approach differs in that we already have the most precise regular expression from the extracted DFA and instead need to produce a simplified yet interpretable regular expression.

\section{Conclusion}\label{sec:conclusion}
Access control policies are reliability-critical artifacts whose correctness depends not on how they were authored but on whether administrators can verify what they actually permit. We argued and empirically showed that LLM-based synthesis cannot solve this verification gap: LLMs can give fluent explanations of policy behavior but cannot reason about the policy semantics with reliability-grade precision. We addressed this gap directly with \tool, a neurosymbolic architecture which pairs symbolic methods with an LLM to generate human-readable characterizations of requests allowed by a policy. \tool ensures the fidelity of the resulting characterization using model counting techniques. Across 746 policies spanning three cloud service providers, \tool produced precise human-readable characterizations with mean similarity 0.93. In a study of 41 practitioners, \tool improved the policy-review fault detection by over 50 percentage points on the hardest subtask. In future work, we intend to extend this neurosymbolic verification pattern to verified policy generation and repair

\section{Artifact and Data Availability}
An anonymized web demo of \tool is available at \url{https://policysummarizer.xyz/}. The demo allows reviewers to paste or upload policy JSON and view the corresponding allowed-resource characterization, LLM-simplified summary, and fidelity score. The open source \tool implementation, datasets, experiment scripts, and generated artifacts are available in an anonymized repository at \url{https://github.com/fractional-distillation/stunning-spork}.

\newcommand{\nonsort}[1]{} \newcommand{\printfirst}[2]{#1}
  \newcommand{\singleletter}[1]{#1} \newcommand{\switchargs}[2]{#2#1}

\end{document}